\documentclass[aps,prb,twocolumn,groupedaddress,showpacs,amsmath,amssymb,amsfonts,10pt]{revtex4-2}

\usepackage[ruled]{algorithm2e}

\usepackage{algpseudocode}

\makeatletter
\def\@algocf@pre@ruled{}
\makeatother
\AtBeginDocument{\pagenumbering{arabic}}
\usepackage{adjustbox}
\usepackage[caption=false]{subfig}
\usepackage{color}
\usepackage{mathrsfs}
\usepackage{xcolor}
\usepackage[titletoc,toc]{appendix}
\usepackage{enumitem}

\usepackage{romannum}

\usepackage[colorlinks,bookmarks=false,urlcolor=blue, citecolor=blue, linkcolor = blue]{hyperref}

\usepackage{tikz}
\usetikzlibrary{positioning,arrows}
\usetikzlibrary{decorations.pathmorphing}
\usetikzlibrary{decorations.markings}
\usetikzlibrary{calc}
\usetikzlibrary{patterns}
\usetikzlibrary{shapes.misc}
\tikzset{
phys/.style={thick, postaction={decorate}, decoration={markings, mark=at position .7 with {\arrow[]{triangle 45}}}},
res/.style={thick, decorate, decoration={snake,segment length=7, amplitude=1.5}},
arrow/.style={thick, draw=white, postaction={decorate}, decoration={markings, mark=at position .6 with {\arrow[black]{triangle 45}}}}
 }
 
\usepackage{hyperref}
\hypersetup{
    colorlinks=true,
    linkcolor=black,
    citecolor=black,
    filecolor=black,
    urlcolor=black
}

\usepackage{etoolbox}
\patchcmd{\thebibliography}{%
  \section*{\refname}%
}{%
  \section*{\refname}%
  \small
}{}{}
\renewcommand{\refname}{References}

\usepackage[absolute]{textpos}
\usepackage[bottom]{footmisc}

\begin{document}

\title{Functional Decomposition and Estimation of Irreversibility\\ 
in Time Series via Machine Learning}

\author{Michele Vodret}
\thanks{Email: mvodret@gmail.com}
\author{Cristiano Pacini}
\author{Christian Bongiorno}
\thanks{Email: christian.bongiorno@centralesupelec.fr}

\affiliation{Université Paris-Saclay, CentraleSupélec,
Laboratoire de Mathématiques et Informatique
pour la Complexité et les Systèmes,
91192 Gif-sur-Yvette, France }

\date{\today}

\begin{abstract}
This work introduces a method to estimate irreversibility in multivariate time series based on the
well-known mapping to a binary classification problem. Our approach utilizes gradient boosting as
a binary classifier, thus providing a model-free, nonlinear, and multivariate analysis while requiring
minimal calibration of the classifier. An additional functionality of the proposed methodology is
to easily dissect the contributions to the irreversibility of subsets of variable interactions, for instance,
those operating at different time scales. The pipeline is divided into three phases: trajectory
encoding, Markovian order identification, and irreversibility estimation via the classifier; the latter could be refined by hypothesis testing and quantification of
variable interactions' contributions to irreversibility. When applied to financial markets, our findings reveal a distinctive shift:
during stable periods, irreversibility is mainly related to short-term patterns, whereas in unstable
periods, these short-term patterns are disrupted, leaving only contributions from stable, long-term
ones. 
\end{abstract}

\maketitle

\section{Introduction}

Time-irreversible dynamics is displayed in systems at all scales, from biological~\cite{mehta2012energetic,barato2014efficiency,barato2015thermodynamic,gnesotto2018broken} to single individuals~\cite{still2012thermodynamics,lynn2021broken,gilson2023entropy,vodret2023cognitive} up to financial ones~\cite{zumbach2009time,xia2014classifying}. 
In full generality, one can associate a value of irreversibility to single trajectories generated by a given dynamic. This is done via the Kullback-Leibler divergence between the forward trajectory's Probability Distribution Function (PDF) and the time-reversed one.

The scientific community has long been interested in refining its toolbox for quantifying irreversibility in time series, given its tight relationship with the notion of dissipated work in thermodynamics~\cite{kawai2007dissipation}. The initial approach focused on simple systems, for which an analytical description of the system's dynamics is available, and was therefore based on model-dependent estimators; in these cases, standard tools from statistical mechanics of out-of-equilibrium systems can be applied successfully~\cite{cocconi2023entropy}. As the focus shifted toward more complex scenarios for which a model is unavailable or too complicated, different model-free methods were proposed to estimate irreversibility directly from data~\cite{kim2020NN,gnesotto2020learning,seif2021machine}. Below, we will discuss briefly these pieces of literature, without the scope of providing a complete analysis. The interested reader can find more details on the dedicated review~\cite{zanin2021algorithmic}.  %

A crucial distinction is whether the system dynamics is between a finite number of states or an infinite number of states~\cite{roldan2014irreversibility}. In the former, simple plug-in-method estimators~\cite{wang2005divergence} can be used if the statistics is sufficient. However, these methods become unfeasible with increasing trajectory length due to a steep rise in the number of probabilities to estimate. The latter case of infinite state-space dynamics is trickier even for short trajectories. Usually one performs first a coarse-graining, also referred to as symbolization~\cite{roldan2014irreversibility}, de-facto, bringing the problem back to one with finite states, i.e., solvable with a plug-in-method estimator. Methods that do not rely on symbolization have been proposed to quantify irreversibility in univariate systems with infinite state-space, such as the use of Markov models~\cite{roldan2014irreversibility} or visibility graph techniques~\cite{lacasa2012time}. 

Recent efforts tackled the infinite state-space case by leveraging Machine-Learning (ML) techniques, mainly focusing on Neural Network (NN) estimators~\cite{kim2020NN,seif2021machine}, but see also~\cite{shang2021directed}. However, one drawback of using NN is that the architecture must be properly tailored to the specific system under analysis; for instance, irreversibility with these tools has been analyzed only for Markovian systems of order one~\cite{kim2020NN,seif2021machine}. The potential structural variations in NNs are virtually limitless, which on the other side, makes the calibration of NNs a complex task. Without a well-designed configuration, issues such as vanishing and exploding gradients may occur~\cite{hanin2018neural}, rendering the NN model unusable. Moreover, calibration demands significant computational resources. Training most architectures on standard laptop CPUs is extremely challenging, and even with the use of GPUs, limitations arise in dedicated memory, GPU costs, and the growing concern over their CO2 emissions impact due to increased power consumption \cite{guan2024reaching}.     

In this work, we provide researchers with an easy-to-use and flexible tool for estimating irreversibility in generic, highly multi-dimensional time series without further constraints.  Our method for irreversibility estimation relies on Gradient Boosting (GB) classifiers. GB classifiers are nowadays one of the most popular and successful ML techniques. Unlike NNs, GB's outcomes are quite robust with respect to changes in the architecture~\cite{bentejac2021comparative} and can run efficiently on standard personal computers, delivering results within a reasonable time frame. Excluding specific tasks like image classification or large language models, which have advanced through extensive global collaboration, GB generally outperforms NN classifiers in most day-to-day scenarios~\cite{mcelfresh2024neural}. GB has outperformed most machine learning methods in classification tasks~\cite{chen2015xgboost,mcelfresh2024neural}; this is interesting to our purposes since the problem of estimating the irreversibility can be recast as a binary classification task. This is illustrated in the following section; first by following common knowledge, and then immediately formalized~\cite{jarzynski2012equalities}. 

The use of GB over NNs offers the advantage of being suitable for hypothesis testing. While both methods can potentially be used for this purpose, the complex calibration of NNs makes them ideal for single-use purposes. On the other hand, the ease and speed of calibration for GB makes it the perfect tool for testing different hypotheses on an unknown system. To this end, this paper focuses on functional decomposition \cite{molnar2020quantifying}, a term popular among engineers and deeply rooted in physical concepts. Functional decomposition involves attempting to break down an unknown function to estimate its non-interacting component to assess its modularity. From a physicist's perspective, this analysis can provide insight into the intricate variable coupling pattern of a multivariate system, allowing for the falsification of models and testing general hypotheses. This type of analysis is particularly important when dealing with complex systems, such as financial markets, where most of the complexity is hidden in intricate interaction patterns~\cite{morel2022scale} that popular econometrics models cannot capture. Fortunately, modern machine learning tools make this type of analysis easily accessible \cite{zupan1997machine,sundararajan2020many}, and we will discuss it thoroughly in this work.

The content of the present paper is organized as follows: 
in section~\ref{sec:1}, we discuss the connection of the statistical mechanics estimation of the irreversibility to the ML binary classification problem.  
In section~\ref{sec:2}, we first discuss generic properties of irreversibility varying the time series length, and then we present an algorithm to estimate irreversibility given a set of trajectories with a generic classifier. Section~\ref{sec:classifiers} is devoted to explaining in depth the advantages of GB by comparing it with simple alternatives.
In section~\ref{benchmarks}, we analyze the functionalities of GB on three simple models for which irreversibility can be analytically characterized. In section~\ref{sec:finance}, we apply the GB binary classifier to estimate irreversibility in a comprehensive financial dataset. In doing so, we show how GB techniques allow decomposing the irreversibility in contributions related to different variable interactions, unveiling insights into the underlying dynamics. Section~\ref{sec:5} concludes.

\section{Irreversibility as a Binary Classification Problem}\label{sec:1}

Suppose you see a movie of a river flowing uphill. Even without knowledge about Newtonian dynamics, experience tells you that it is a zero-probability event. In other words, since you have always seen water flowing downhill, you are sure the movie is playing backward. If the dynamics is stochastic, as in mesoscopic biological and financial systems, distinguishing forward from backward dynamics depends on how much the trajectory is irreversible and intuitively relies on a signal-to-noise ratio. In summary, given an initial knowledge of the forward dynamics, the more it breaks temporal reversal symmetry, the easier it is to classify forward and backward trajectories. 

Consider now a balanced training set where half of the trajectories come from the forward dynamics and the remaining half are time-reversed. We will denote an element of the first group as $\overrightarrow{\boldsymbol{q}} = \{\boldsymbol{q}_{1}, \cdots, \boldsymbol{q}_{t}\}$ and an element of the second as $\overleftarrow{\boldsymbol{q}}=\{\boldsymbol{q}_{t}, \cdots, \boldsymbol{q}_{1}\}$. We denote with $\mathcal{P}(\overrightarrow{\boldsymbol{q}}|F)$ the probability to generate the trajectory $\boldsymbol{q}$ when performing the forward process $F$, analogously to what is done in  Ref.~\cite{jarzynski2012equalities}; similarly, $\mathcal{P}(\overleftarrow{\boldsymbol{q}}|F)$ is the probability to generate the time-reversed trajectory performing the forward process.

Time irreversibility of the trajectory $\boldsymbol{q}$ is defined in terms of the Kulback-Leibler divergence between the probability distributions introduced above, i.e., 
\begin{align}
    \label{KL_def}
        D[\mathcal{P}(\overrightarrow{\boldsymbol{q}}|F),\mathcal{P}(\overleftarrow{\boldsymbol{q}}|F)] :=
        \left\langle \log \left( \frac{\mathcal{P}(\overrightarrow{\boldsymbol{q}}|F)}{\mathcal{P}(\overleftarrow{\boldsymbol{q}}|F)}\right)\right\rangle_{\mathcal{P}(\overrightarrow{\boldsymbol{q}}|F)} &
        \\
        \label{formula}
        =\left\langle \log \left( \frac{\mathcal{P}(F|\overrightarrow{\boldsymbol{q}})}{\mathcal{P}(F|\overleftarrow{\boldsymbol{q}})}\right)\right\rangle_{\mathcal{P}(\overrightarrow{\boldsymbol{q}}|F)} &,
\end{align}
where the second equation is obtained by invoking Bayes' theorem~\cite{jarzynski2012equalities} together with the symmetry of our balanced training set. Crucially, 
Eq.~\eqref{formula} states that one can calculate the irreversibility by computing first the conditional probabilities $\mathcal{P}(F|\overrightarrow{\boldsymbol{q}})$ and $\mathcal{P}(F|\overleftarrow{\boldsymbol{q}})$ and then averaging the logarithm of their ratio on $\mathcal{P}(\overrightarrow{\boldsymbol{q}}|F)$. 

This result opens the door to the use of ML architectures to estimate irreversibility in generic time series, as they can be used as supervised binary classifiers to obtain $\mathcal{P}(F| \boldsymbol{q})$. Note that this approach does not rely on constructing a tailored functional to be minimized~\cite{kim2020NN} and has the advantage of being fully consistent with standard stochastic thermodynamics principles.
To show this, let us first rewrite $\mathcal{P}(F|\boldsymbol{q})$ as
\begin{equation}
\label{logistic}
    \mathcal{P}(F|\boldsymbol{q}) = \frac{1}{1+e^{-\ell(\boldsymbol{q})}},
\end{equation}
where we have simply implicitly defined the logit $\ell({\boldsymbol{q}})$. Finally the well-known Kawai-Parrondo-Van den Broeck theorem~\cite{kawai2007dissipation} follows by inserting Eq.~\eqref{logistic} in the r.h.s. of Eq.~\eqref{formula} together with the following relation obtained from trivial symmetries $\mathcal{P}(F|\overleftarrow{\boldsymbol{q}}) = 1-\mathcal{P}(F|\overrightarrow{\boldsymbol{q}})$:
\begin{equation}
\label{eq:logit}
    D[\mathcal{P}(\overrightarrow{\boldsymbol{q}}|F), \mathcal{P}(\overleftarrow{\boldsymbol{q}}|F)] = \langle \ell(\overrightarrow{\boldsymbol{q}})\rangle_{\mathcal{P}(\overrightarrow{\boldsymbol{q}}|F)}.
\end{equation}
In physical language, the equation above states that irreversibility in stationary dynamics is equal to the average stochastic dissipated work over the trajectories and constitutes a refinement of the second law of thermodynamics. 
Based on Eq.~\eqref{eq:logit} we can rephrase the distinction between model-dependent and model-free approaches to quantify irreversibility:
the functional form of $\ell(\boldsymbol{q})$ is known in the former, while it is not in the latter and it can be estimated using ML methods.

\section{Irreversibility Estimation Framework}
\label{sec:2}

\subsection{Trajectory 
\label{sec:length}
Length and Information Content} 
It is clear that the longer the trajectories of an irreversible dynamics, the simpler it is to distinguish forward from backward trajectories, implying a higher irreversibility. However, we shall see in what follows that a quantitative increase of irreversibility for long trajectories might not be associated with the discovery of new irreversible mechanisms.

To focus on the irreversibility contributions appearing at different timescales, it is useful to define the irreversibility associated with slices of trajectories with length $t$ as $D^{(t)}$.

In the simplest case of trajectories with independent increments, irreversibility increases with a constant rate, i.e., $D^{(t+1)}- D^{(t)} = \text{const.}$ for all $t$.  In fact, in this case, $\mathcal{P}(\boldsymbol{q}|F)$ can be written in terms of a product of identical probabilities of one-step increments.

In the case of a Markovian process of order $\tau$, the above condition is valid only for $t\geq \tau$~\cite{rached2004kullback,roldan2010estimating}, i.e., 
\begin{equation}
\label{eq:KL_mark}
    D^{(\tau+h+1)}- D^{(\tau+h)} = \text{const.}, \ \text{for} \ h\geq 0.
\end{equation}
This means that no new mechanisms inducing irreversibility are expected when studying trajectories of length $t\geq \tau$.

These considerations should have convinced the reader that if some knowledge about the Markovianity of the process at hand is known, one could leverage it when studying irreversibility. Trivially, nothing qualitatively new is learned from the analysis of trajectories of length $t>\tau$. On the other hand, when the underlying model is not known, one can have an estimate of $\tau$ if the condition defined in Eq.~\eqref{eq:KL_mark} is satisfied by the sample estimate of $D^{(t)}$. 

\subsection{Trajectory Encoding}
\label{sec:encoding}
A necessary step before applying any classifier is to choose an encoding of the trajectory. In general, we can think of a trajectory as a sequence of $t$ positions in the $m$-dimensional space, i.e.,  $\boldsymbol{q}=\{\boldsymbol{q}_1, \cdots, \boldsymbol{q}_t\}$ where $\boldsymbol{q}_i= \{ q_{i1},\cdots, q_{im}\}$ is a vector of dimension $m$.
An encoding will reduce the dimensionality of the problem from $t\times m$ to $k$ features $\boldsymbol{x}=\{x_{1},\cdots, x_{k}\}$, which can be real numbers, discrete numbers, or even symbols.  

The naive encoding in the case of $m=1$ is the direct mapping  $q_i = x_i$ or, for $m>1$, a flattened version of $\boldsymbol{x}$, i.e., $q_{\left(i-1\right) m +j}=x_{ij}$. In these cases, the encoding procedure is a bijective transformation, and therefore no information is lost. This method could raise some problems in the estimation of the classifier's probability function $\mathcal{P}(F|\boldsymbol{q})$. For example, let us imagine having a Brownian walker $\boldsymbol{q}$, i.e., a process with independent increments. If naive encoding is employed, very few statistics are recorded for each position, leading to poor irreversibility estimates. A way out is to consider an alternative encoding given by increments, i.e., $\overrightarrow{x}_i = q_i - q_{i-1}$; note that this encoding does not discard irreversibility-related information. 

Let us explicitly note that according to our formalism, the time inversion is applied to $\boldsymbol{q}$, not to the encoding $\boldsymbol{x}$; this means that $\overleftarrow{\boldsymbol{x}}$ does not always translate into a simple inversion of the indices of $\overrightarrow{\boldsymbol{x}}$. For example, in the case of the first differences $\overrightarrow{x}_i=q_{i}-q_{i-1}$ discussed above, a time reversion on $\overrightarrow{\boldsymbol{x}}$ will imply both an inversion of the indices and a change of the sign. 

The logit itself $\ell(\boldsymbol{q})$ can be viewed as a non-linear coupled encoding of the trajectory into a single real value that contains all the information about the system's irreversibility. However, this encoding is available only for systems with known and analytically solvable dynamics. We will see in Sec.~\ref{1dim} that for the case of the Brownian walker $\ell(\boldsymbol{x}) \propto \boldsymbol{x}$.

In the general case, given an encoding and the knowledge of the true probabilities function, the following inequality holds:
\begin{equation}
\label{eq:ineq}
    D[\mathcal{P}(\overrightarrow{\boldsymbol{x}}|F), \mathcal{P}(\overleftarrow{\boldsymbol{x}}|F)] \leq D[\mathcal{P}(\overrightarrow{\boldsymbol{q}}|F), \mathcal{P}(\overleftarrow{\boldsymbol{q}}|F)],
\end{equation}
where the equality is reached in the cases where the encoding does not discard irreversibility-related information content.
However, since $\mathcal{P}(\boldsymbol{q}|F)$ and $\mathcal{P}(\boldsymbol{x}|F)$ must be estimated from finite sample size data, Eq.~\eqref{eq:ineq} will not be necessarily satisfied from sample estimates. 

A key observation that makes it worth spending some time on the encoding is that one can use it to test specific model dependencies; it is not just a matter of model performance. In fact, if the measured irreversibility among two different encodings does not change, then one can conclude that system irreversibility does not come from those differences.

\subsection{Irreversibility Estimation from Finite Samples}
\label{sec:pseudo}

This section specifies the procedure for training a classifier to estimate irreversibility. In many practical cases, multiple trajectories are not available. However, if the trajectory is sufficiently long and stationary, one can study the irreversibility of the original trajectory's slices of length $t$.

First, given a set of $N_{\text{train}}$ encoded stationary trajectories of length $k$ one constructs the reversed paths and labels them accordingly. Doing so, one obtains a balanced set of dimension $2 N_{\text{train}}$ of encoded trajectories half labeled by $F$ (or 1) and half labeled by $B$ (or 0). 

The classifier architecture is trained on the forward and backward encoded trajectories to predict the classes $F$ or $B$. The training phase gives as an output a model-free classifier's probability function $\hat{\mathcal{P}}(F|\boldsymbol{x})$.   
Note that using $\hat{\mathcal{P}}(F|\boldsymbol{x})$ to compute Eq.~\eqref{formula} on the same set of encoded. 
Such an issue is particularly relevant when the number of parameters of the chosen classifier model is high.

To overcome this issue, a Cross-Validation (CV) approach might be used, de-facto computing the irreversibility on another set of $2 N_{\text{test}}$ encoded trajectories.
In practice, given a set of $N$ encoded trajectories, splitting them into a train and test (usually $80\%-20\%$) will imply computing the irreversibility on a small subset of the available data. 

To surmount this second limitation, a $K$-fold CV can be applied. Essentially, one splits the ensemble of $N$ encoded trajectories in $K$ non-overlapping folds. For each partitioning, one constructs the ensemble of reverse encoded trajectories, and finally, the union of $K-1$ folds is used for training, and the remaining fold is used for testing.  
In doing this, one calibrates $K$ models, one for each fold, and then uses them to compute the actual irreversibility from Eq.~\eqref{formula} using the sample mean over the related test set. Finally, one averages the estimates over folds, thereby computing the right-hand side in the final equation of Eq.~\eqref{formula}. 

Trivially, if one is interested in computing a stratified mean of Eq.~\eqref{formula} according to some trajectory's metadata, the individual contribution of the trajectories to the irreversibility can be averaged at the end. 

The general algorithm to obtain a sample estimate of irreversibility $\hat{D}^{(t)}$ from slices of length $t$ of the original trajectory is described in Algorithm~\ref{ref:algo}.

\begin{algorithm}[tbh]
\caption{Irreversibility Estimation from a Classifier}
\label{ref:algo}
\hrulefill
\vspace{0.1cm}
\begin{algorithmic}[0]

\Require $\{\overrightarrow{\boldsymbol{q}_h}\}_{h=1}^N  (N \ \text{trajectories} \ t \times m)$
\Ensure $\hat{D}^{(t)}$ (estimated irreversibility)
\State $\{\overrightarrow{\boldsymbol{x}_h}\}_{h=1}^N \leftarrow \{\overrightarrow{\boldsymbol{q}_h}\}_{h=1}^N$ (choose encoding in $k$ features)
\State $\overrightarrow{\mathcal{S}}_{1}, \ldots, \overrightarrow{\mathcal{S}}_{K} \leftarrow  \{\overrightarrow{\boldsymbol{x}_h}\}_{h=1}^{N/K}, \dots, \{\overrightarrow{\boldsymbol{x}_h}\}_{h=N-N/K+1}^{N}$~(K-Fold)

\State  $\overleftarrow{\mathcal{S}}_{1}, \ldots, \overleftarrow{\mathcal{S}}_{K} \leftarrow \text{reverse}(\overrightarrow{\mathcal{S}}_{1}, \ldots, \overrightarrow{\mathcal{S}}_{K}$)
\State Label $\{\overrightarrow{\mathcal{S}}_i\}_{i=1}^K \ $ with $1 \ $ and $\ \{\overleftarrow{\mathcal{S}_i}\}_{i=1}^K $ with $ \ 0$

\hspace{-0.65cm}\For{{$i = 1 $ \text{to} $ K$}}{
    \State Train classifier on $\bigcup_{j \neq i} \left\{ \overrightarrow{\mathcal{S}}_{j} \bigcup \overleftarrow{\mathcal{S}}_{j} \right\}$ 
    \State $\{\overleftarrow{p}_h\}_{h=1}^{N/K} \gets  \{ \hat{\mathcal{P}}(1| \boldsymbol{x} )\  \text{for}\  \boldsymbol{x} \ \text{in} \ \overrightarrow{\mathcal{S}}_{i} \}$
    \State $\{\overrightarrow{p}_h\}_{h=1}^{N/K} \gets  \{ \hat{\mathcal{P}}(1| \boldsymbol{x} )\  \text{for}\  \boldsymbol{x} \ \text{in} \ \overleftarrow{\mathcal{S}}_{i} \}$
    \State $\hat{D}_i^{(t)} \gets  \cfrac{K}{N}\sum_{h=1}^{N/K} \left[  \log (\overrightarrow{p}_h ) - \log ( \overleftarrow{p}_h) \right]$}

\State $\hat{D}^{(t)} \gets \frac{1}{K} \sum_{i=1}^{K} \hat{D}^{(t)}_i$
\State \textbf{return } 0 
\end{algorithmic}
\hrulefill
\end{algorithm}

\section{Three architectures for binary classifiers}
\label{sec:classifiers}
\subsection{Logistic Regression}
\label{LogReg}
The simplest approach to binary classification is the Logistic Regression (LR), i.e. a linear regression on the logit $\ell(\boldsymbol{x})$ appearing in Eq.~\eqref{eq:logit}, i.e.,
\begin{equation}\label{eq:LR}
    \hat{\ell}(\boldsymbol{x}) = \hat\alpha + \hat{\beta}_1 x_1 + \hat{\beta}_2 x_2 + \cdots + \hat{\beta}_k x_k. 
\end{equation}
The parameters $(\hat\alpha, \hat{\boldsymbol{\beta}})$  are chosen to minimize the log-loss, also known as binary cross-entropy. 

Given the scenario we are looking at, where initial data are labeled as forward and we generate artificially backward ones, the $\alpha$ parameter in Eq.~\eqref{eq:LR} is zero. This means that a priori, a given sample in the train set has the same chance of being forward or backward.

By construction, standard LR can correctly handle binary classification tasks for which $\ell(\boldsymbol{x})$ is linear in the features. Examples are given by cases where the two conditional probability distributions belong to the same exponential family, with equal dispersion parameter~\cite{jordan1995logistic}. 

If analytical formulas for $\ell(\boldsymbol{x})$ are available, one can leverage them and successfully apply LR. Indeed, in the general case, $\ell(\boldsymbol{x})$ will be given as a combination of features' interaction terms. If these are provided to the LR, then it can be successfully applied. In such contexts, the LR should be preferred to more complex models, which will be prone to overfitting. 

\subsection{Naive Coarse-graining}\label{sec:coarse}
When one cannot make assumptions about the underlying model $\ell({\boldsymbol{x}})$, a naive model-free approach to estimate irreversibility may rely on uniform phase space partitioning of ${\boldsymbol{x}}$. In this case, one approximates $\mathcal{P}(F|{\boldsymbol{x}})$ as a histogram obtained by counting the occurrences of ${\boldsymbol{x}}$ in the forward and backward batches. Finally, one obtains the irreversibility by plugging them directly into Eq.~\eqref{KL_def}. This method is similar to the plug-in method combined with symbolization,~\cite{roldan2014irreversibility}, although here it is applied on $\mathcal{P}(F|\boldsymbol{x})$ and not on $\mathcal{P}(\boldsymbol{x}|F)$, and in fact shares with it some relevant issues.  

For instance, the partition width should be adapted to the sample size. A too-coarse partition for the sample size might fail to detect the irreversibility, leading to $\hat{\mathcal{P}}(F|{\boldsymbol{x}})=0.5$ in the limit where the partition is given by only one bin. On the other hand, a too-fine coarse-graining can rapidly lead to problems related to extreme probability estimates in small sample bins, which makes the log ratio in Eq.~\eqref{formula} over-estimated or undefined.  As a rule of thumb, the larger the sample size, the finer the partition can be. However, finding the right balance is hard and, in any case, might not be efficient. 
Ad-hoc assumptions such as biasing the empirical distributions or neglecting contribution from bins where the log ratio in Eq.~\eqref{formula} is undefined are an attempt to address the extreme probability estimates. 

Below, we explore an alternative approach based on non-homogeneous partitioning. This is useful since while high-density regions might allow for finer partitioning, the bins must be larger on the low-density ones. 

\subsection{Gradient Boosting}\label{sec:GB}

GB is a tree-based technique that can capture non-linearity and feature interactions hidden in a multivariate dataset. GB can be applied for regression or classification problems, the latter being the functionality we are interested in. When applied to quantify irreversibility in time series, the GB classifier combines efficient non-homogeneous coarse-grainings of the trajectory space allowing the estimate $\mathcal{P}(F|{\boldsymbol{x}})$ and, therefore, the irreversibility via Eq.~\eqref{formula}. We explain below in detail how GB works.  

The elementary unit of a tree-based classifier is the decision tree $\Theta(\boldsymbol{x}): \boldsymbol{x} \rightarrow \mathbb{R}$. 
A tree is organized into a set of nodes linked together, representing a sequence of if-else conditions, and leaves, representing the contribution to the logit value obtained at the end of each sequence of nodes.   

Each node is associated with a splitting of the feature space in an axis-parallel way, e.g.~$x_{i}>\text{const}$. These splits effectively partition the feature space to misclassify the least data points.  Splits in the nodes are chosen iteratively to minimize a chosen loss function, typically the log-loss in classification tasks.  The maximal number of subsequent tree nodes is known as depth $d$, and choosing it properly is crucial to overcome over-fitting. In fact, while one could build a single tree with a large depth, it is well-known that more efficient alternatives are available~\cite{xgboost}. These include ensembling methods such as Random Forest and GB. 

\subsubsection{Boosting}
\label{boosting}
The idea behind GB is to build shallow trees $\Theta({\boldsymbol{x}})$, i.e., trees with a fixed, ideally small, maximal depth, and combine them iteratively into a more robust learner. In formulas:  
\begin{equation}
    \hat{\ell}({\boldsymbol{x}}) = \hat\ell_{0}({\boldsymbol{x}})+ \lambda \sum_{i=1}^{N_\text{trees}} \hat{\Theta}_i({\boldsymbol{x}}),
\end{equation}
the initial predicted logit $\hat\ell_0({\boldsymbol{x}})$ is only related to the fraction of the two classes that, in our case is $0.5$ by construction; therefore, $\hat\ell_0({\boldsymbol{x}}) = 0$.  Each calibrated tree  $\hat{\Theta}_i({\boldsymbol{x}})$ refines the prediction, and it is scaled by factor $\lambda \ll 1$ known as the learning rate, in line with what is done in gradient-based techniques. The learning rate, the number of trees $N_{\text{trees}}$, and the maximal depth are crucial parameters of the GB method. An ideal robust learner should have a small $\lambda$, a large $N_\text{trees}$, and a small maximal depth. 

The prediction of each tree is refined iteratively such that 
 \begin{equation}
    \hat{\ell}_{i}({\boldsymbol{x}}) = \hat{\ell}_{i-1}({\boldsymbol{x}})+ \lambda \hat\Theta_i({\boldsymbol{x}}),
\end{equation}
where $\hat\Theta_i({\boldsymbol{x}})$  fits the pseudo-residuals $\hat{\epsilon}_i({\boldsymbol{x}})$ of the previous models
\begin{equation}
    \hat{\epsilon}_{i}(\boldsymbol{x}) = \begin{cases} 1 - \hat{\mathcal{P}}_{i-1}(F|\boldsymbol{x}), \ \text{if} \ \boldsymbol{x} \ \in \ F
    \\
    -\hat{\mathcal{P}}_{i-1}(F|\boldsymbol{x}), \quad \  \text{if} \ \boldsymbol{x} \ \not\in \ F,
    \end{cases}
\end{equation}
where $\hat{\mathcal{P}}_{i}(F|\boldsymbol{x})$, analogously to  Eq.~\eqref{logistic}, is the sample-based conditional probability associated with logit  $\hat{\ell}_{i}(\boldsymbol{x})$. Here we use the convention that when $i=N_{\text{trees}}$
the subscript can be omitted both in the logit and the probability. 

It is important to note that every iteration will lower the residue, leading for large $N_{\text{tree}}$ to overfitting.  To prevent this, usually, the loss on a validation dataset is monitored during the training phase, and the iteration is stopped when the loss does not decrease on the validation dataset for ten iterations; this procedure is known as the early stopping rule.

\begin{figure*}[t!]
\centering

    \centering
\includegraphics[scale=0.3]{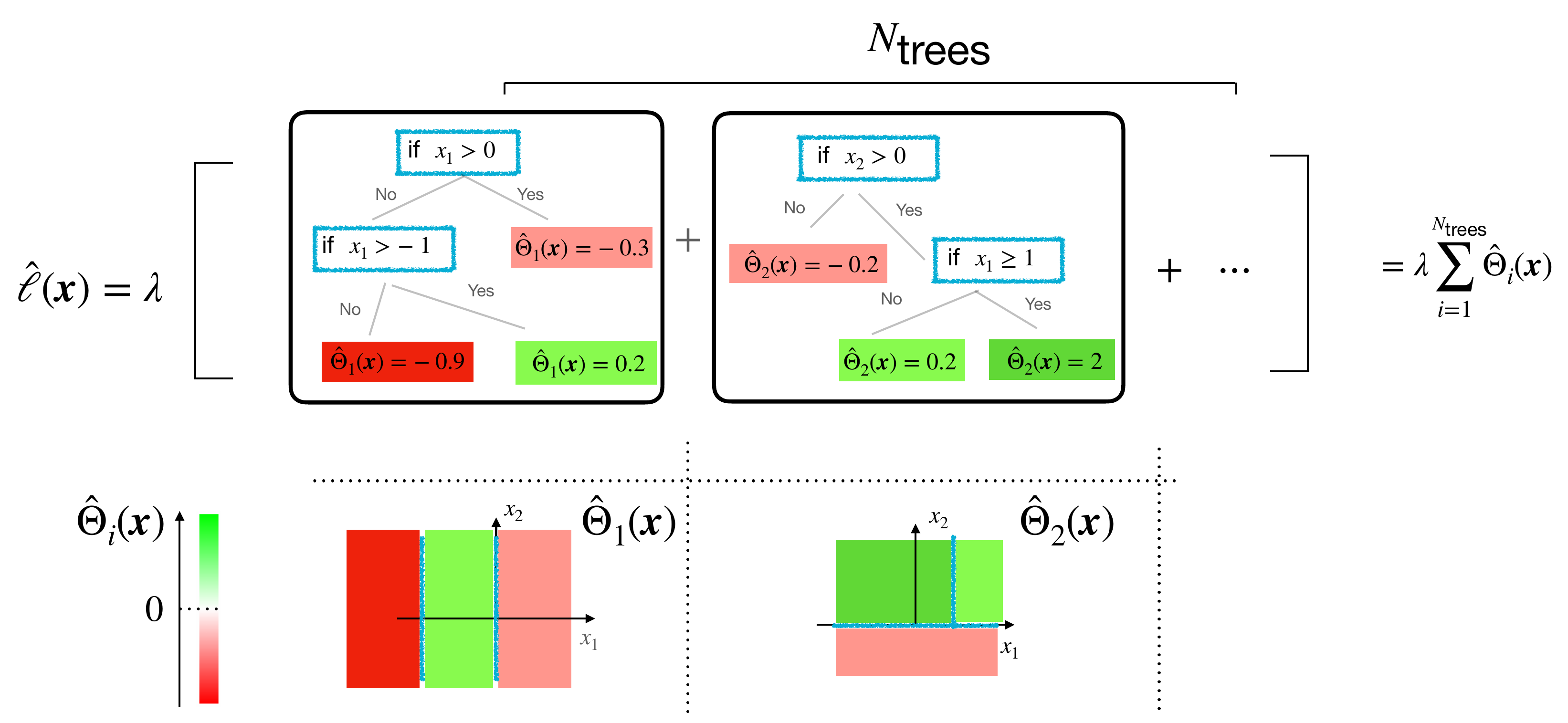}

\caption{A visual example of how GB in binary classification works with bi-dimensional input feature space. This image refers to a trained GB with a given set of splits and to a sample given by ${\boldsymbol{x}}=\{x_{1}, x_{2}\}= \{-2,1\}$. On top, it shows how the logit $\hat{\ell}(\boldsymbol{x})$ is computed, with contributions $\hat{\Theta}_i(\boldsymbol{x})$ with $i \in \{1,2, \dots, N_{\text{trees}}\}$ coming from $N_\text{trees}$ different trees. The bottom part of the figure shows graphically how the split in each tree's node partitions the input feature space. Note that while the first tree models a non-linearity on $x_1$, the second tree models an interaction between $x_1$ and $x_2$. }
\label{fig:xgboost}
\end{figure*}

\subsubsection{Interactions}\label{sec:interaction}

Non-linearity on a single feature $x_{i}$ can be accounted for by boosting single node trees because the elementary operation, i.e., the splitting on each node of each decision tree is itself a non-linear operation.  However, boosting single-node trees is not enough to capture coupling between different features. To capture a coupling between $n$ features, one needs trees with a depth $ d \gtrsim n$. Although $d = n$ should be enough in this case, in practice, it is convenient to choose $d$ slightly larger than $n$ to help the optimization of the shallow trees in the possibly highly dimensional interaction space. The key characteristic of a well-constructed GB classifier is to be parsimonious on the depth so that the tree can capture the maximal coupling between the input features while boosting takes care of non-linearities.

Let us consider an example where $d = 2$. We will refer to Fig.~\ref{fig:xgboost}. The logit of an observation ${\boldsymbol{x}}$ is computed by combining the $N_{\text{trees}}$ of the figure; those explicitly represented have a depth $d = 2$. 
Suppose that the observation is ${\boldsymbol{x}}=\{ x_1, x_2\} = \{-2,1\}$. Once passed to the first tree, it follows the first split, taking the left branch and ending in the bottom left leaf node, yielding $\hat{\ell}_1(\boldsymbol{x}) = -0.9$. Instead, following the split of the second tree it ends in the leaf node with output $\hat{\ell}_2(\boldsymbol{x}) = 0.2$. Note that in the first tree, the classification only depends on feature $x_1$; therefore, this tree could be decomposed into two trees with a depth of one. Instead, the second tree splits on both $x_1$ and $x_2$ in the same branch and cannot be decomposed similarly. When a tree splits along two features within the same branch, it can effectively capture the joint effect of  $x_1$ and $x_2$; for instance, it can capture a multiplicative interaction.
The discussion above should have convinced the reader that the maximum depth parameter limits the maximum interaction order that the classifier can capture.

On top of the global constraint induced by the maximal depth of the shallow decision trees, the XGBoost package~\cite{xgboost} allows us to consider more specific constraints via the so-called interaction constraints (ICs). 
Specifically, by controlling which combination of features can belong to the same branch, one can control the specific interactions the model is able to discover. Suppose that the dimensionality of the dataset is 5. An example of IC can be $\{\{x_1,x_2\}, \{x_3, x_4, x_5\}\}$, where each inner list is a subset of features that are allowed to interact with each other in a given tree's branch.

A Python code implementing the XGBoost irreversibility estimator is available on PyPI~\cite{irreversibility_estimator_2024}.

\begin{table*}[tbh]
    \centering

    \renewcommand{\arraystretch}{1.5} 
        \begin{tabular}{cccccccc}
        \hline
        \vspace{0.2cm}
        \textbf{} & \textbf{} & Encoding ($\overrightarrow{\boldsymbol{x}}$)  &   Coupling  & Linearity & \textbf{$\mathcal{P}(F|x)$} & \textbf{$D$} \\
        \vspace{0.2cm}
        $a)$ & \ Gaussian increments \ & $ \{q_t-q_{t-1}\}$ &  no     & yes & $\left[1+\exp\left(-\cfrac{2\mu x}{\sigma^2}\right)\right]^{-1}$ & $ 2\cfrac{\mu^2}{\sigma^2}$ \\
         \vspace{0.2cm}
        $b)$ & \ Cauchy increments \  & $\{q_t-q_{t-1}\} $ &  no  & no & $\cfrac{1}{2}+\cfrac{x \xi}{\gamma^2 + \xi^2 + x^2}$& $\log\left( 1+\cfrac{\xi^2}{\gamma^2}\right)$\\
        $c)$ & \ Brownian gyrator \  &  $\{q_{t1},q_{t2},q_{t+1,1},q_{t+1,2} \}$  &   yes  & yes & $\left[1+\exp\left(-2A\cfrac{x_1x_3-x_2x_4}{2B-A^2-B^2}\right)\right]^{-1}$  & $\quad \cfrac{4A^2}{2B-A^2-B^2} $ \vspace{0.2cm}\\   
        \hline
        
    \end{tabular}
    \caption{Summary table for the simple cases analyzed. }
    \label{tab:formulas}
\end{table*}

\section{Simple case studies}
\label{benchmarks}
In this section, we explore how crucial is the choice of the classifier's architecture for a correct irreversibility estimation when trajectories from models with different complexities are given. 
 
We consider two one-dimensional stochastic processes with independent increments given respectively by $a)$ Gaussian distribution and $b)$ Cauchy distribution.  
Then, we consider $c)$ the Brownian gyrator model, i.e., a two-dimensional, stationary, Gaussian and Markovian process of order one.
Table~\ref{tab:formulas} summarizes the properties and results of these three simple models.

\subsection{Linear 1-dimensional case}
\label{1dim}
First, we consider a random walker 
\begin{equation}\label{eq:driftwalk}
    q_{t} = q_{t-1} + \eta_{t},
\end{equation}
with independent Gaussian increments $\eta_t$ given by 
\begin{equation}
\label{eq:gaussian}
\mathcal{P}(\eta_t; \mu, \sigma) = \frac{1}{\sqrt{2\pi\sigma^2}} \exp\left(\cfrac{(\mu-\eta_t)^2}{2\sigma^2}\right).
\end{equation}

Given that increments are independent, the most natural encoding is the singlet $\overrightarrow{x} = \{q_{t}-q_{t-1}\}$, so that $\overleftarrow{x}=-\overrightarrow{x}$. Therefore, the forward probability of the encoded variable is $\mathcal{P}(x|F) = \mathcal{P}(x; \mu, \sigma)$, while the probability distribution for the backward jump has mean $-\mu$.  So the classification task reduces in distinguishing between two Gaussian distributions with the same $\sigma$ and opposite $\mu$.

The conditional probability $P(F|x)$ that $x$ comes from the forward distribution can be derived from Bayes' theorem and has a logistic shape with
\begin{equation}
\label{eq:logit_gauss}
    \ell(x) = \cfrac{2\mu x}{\sigma^2}.
\end{equation} 
In this case, the logit (defined in Eq.~\eqref{logistic}) is linear in the encoded variable.
Therefore, the LR provides the correct ansatz (see Eq.~\eqref{eq:LR}), and it reduces to estimate a single slope of $2 \mu/\sigma^2$. In this scenario, using GB is overly complicated. Although a calibration with trees of depth one yields comparable results on a sufficiently large dataset, the linearity of Eq.~\eqref{eq:logit_gauss} is modeled with GB as a combination of piece-wise constant elements of $x$. This is, of course, not ideal and unnecessary in this case.

The irreversibility is the average of the logit on $\mathcal{P}(x|F)$ (Eq.~\eqref{eq:logit}) so that $D$ is proportional to the square of the signal-to-noise ratio $\mu/\sigma$ (see formulas reported in Table~\ref{tab:formulas}).

\begin{figure}[tbh]
\includegraphics[width = 0.8\columnwidth]{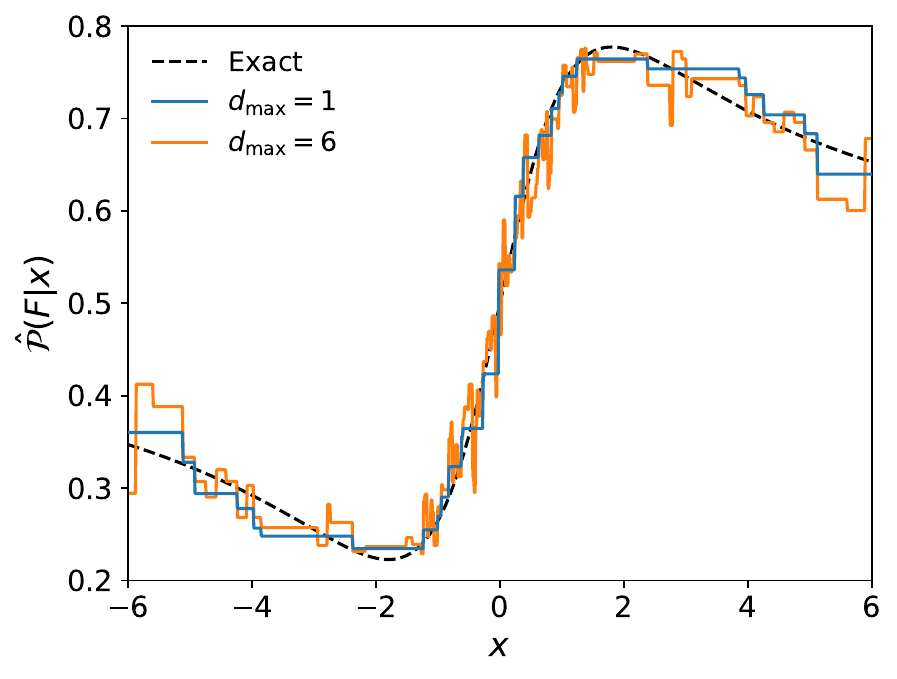}
\includegraphics[width = 0.8\columnwidth]{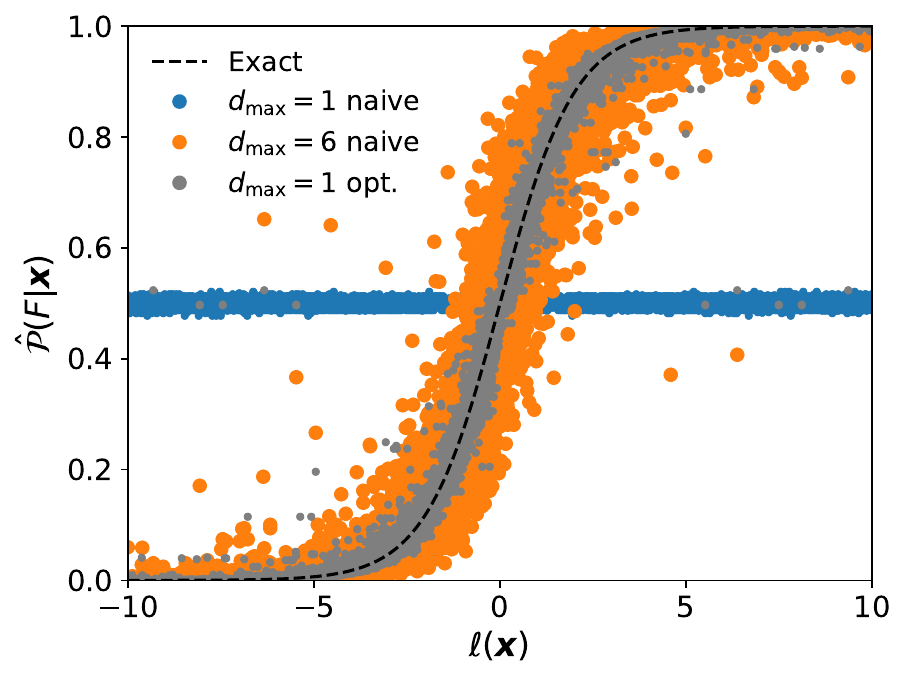}
\caption{
The upper panel shows the estimated $\hat{\mathcal{P}}(F|x)$ as a function of $x$ for the GB classifier in the non-linear 1-dimensional case study, i.e., the random walker with Cauchy distributed increments. The variable $x$ is extracted from a Cauchy with parameters  $\xi=1$ and $\gamma=1.5$. GB is calibrated from $N=10,000$ trajectory using a single-round CV with $80\%-20\%$ for train and test, respectively. The figure reports two different calibrations of GB, i.e., with a maximum depth $d_{\text{max}}$ of $1$ and $6$ and the exact analytical solution. The lower panel shows the estimated $\hat{\mathcal{P}}(F|\boldsymbol{x})$ from GB as a function of theoretical optimal logit $\ell(\boldsymbol{x})$ in the Brownian gyrator case, for $A=0.3$ and $K=0.1$ with $N=100,000$ trajectory using a single-round CV with $80\%-20\%$ for train and test. The points are obtained from the test set. The figure shows exact analytical expectations, two models calibrated with the naive encoding of the position and different maximal depth constraints, and a model calibrated with the optimal encoding of Eq.~\ref{eq:optimalenc} and a maximum depth of 1.}
\label{fig:gauss_cauchy}
\end{figure}

\subsection{Non-linear $1$-dimensional case}

Now we consider a random walker with Cauchy distributed increments, i.e., 
\begin{equation}
\label{eq:cauchy}
    \mathcal{P}(\eta_t; \xi, \gamma) = \cfrac{1}{\pi \gamma \left[1 + \left(\cfrac{\eta_t - \xi}{\gamma}\right)^2\right]}.
\end{equation}
where the location and the scale parameters are respectively given by $\xi$ and $\gamma$. Given that increments are again independent, the best encoding is the same as the previous case.

Note that the Cauchy distribution does not belong to the exponential family since its moments are not well-defined. This implies that the optimal classifier will not have a logistic shape with linear logit~\cite{jordan1995logistic}. In fact, from simple calculations similar to those of the previous section, one derives
\begin{equation}
    \ell(x) = \log \left(  \pi \gamma \left( \frac{x-\xi}{\gamma}\right) -1 \right).
\end{equation}

$\mathcal{P}(F|x)$ is a non-monotonous function (see the black dashed line in the upper panel of Fig.~\ref{fig:gauss_cauchy}) that tends to $1/2$ as $x$ tends to $\pm \infty$, meaning that big jumps cannot be well classified.

The LR estimator gives a completely off estimate of $\mathcal{P}(F|x)$ leading to a $\beta=0$, which means $\hat{\mathcal{P}}(F|x)=0.5$ for every $x$. Setting a cutoff on the extreme value ($|x|<c$ with $c$ finite), we can obtain at least a correct slope sign even though it is still off quantitatively. On the other hand, GB predicts a correct non-monotonous likelihood by using a combination of piece-wise constant elements of $x$, as shown in the upper panel of Fig.~\ref{fig:gauss_cauchy}.  An interesting observation discussed in Sec.~\ref{sec:coarse} is that the partitions along \(x\) are not homogeneous; these are finer around \(x=0\) where there is more statistical data and coarser for larger \(|x|\). 

Furthermore, it is worth mentioning that a maximum depth of \(d=1\) is sufficient for this type of problem and is less noisy than a more complex alternative with \(d=6\). This is because only single-variable non-linearity needs to be handled in this case, and as explained in Section~\ref{sec:interaction}, using boosted single-split trees is adequate for this purpose.

The formula reported in Table~\ref{tab:formulas} for the irreversibility can be derived from available results in the literature~\cite{chyzak2019closed}.

\subsection{Coupled 2-dimensional case}
\label{MultiDim}

Now, we consider a case that needs a GB architecture of maximum depth $d \geq 2$, i.e., a Brownian particle moving in a 2-dimensional space and subject to a constant radial conservative force and a constant non-conservative (rotational) component, also known as Brownian gyrator. The non-conservative force leads in the stationary state to non-vanishing probability currents~\cite{tome2006entropy}, a clear signature of time-reversal-asymmetry. 

The discrete-time description is given by:
\begin{align}
\label{eq:Brownian_Gyrator}
q_{t+1,1} &= (1-B) q_{t1} + A q_{t2} + \eta_{t 1} \nonumber
\\
q_{t+1,2} &= (1-B) q_{t2} - A q_{t1} + \eta_{t 2} 
\end{align}
where $\eta_{t1}$ and $\eta_{t2}$ are independent centered Gaussian noises with equal variances and $A,B>0$; their variances are chosen so as to have a stationary PDF of the $\boldsymbol{q}$ process given by a two-dimensional Gaussian with unit variance, i.e., $\langle \eta^2\rangle = 2B-B^2 -A^2$. Note that this choice does not affect the irreversibility due to its scale invariance property, which can be derived from the more general invariance under parameter transformations~\cite{lexa2004useful}. 
  
Being Markovian and stationary, we consider the naive encoding given by 
\begin{equation}\label{eq:naiveenc}
   \overrightarrow{\boldsymbol{x}} =\{\overrightarrow{x}_1,\overrightarrow{x}_2,\overrightarrow{x}_3,\overrightarrow{x}_4\} =\{q_{t1}, q_{t2}, q_{t+1,1}, q_{t+1,2}\}. 
\end{equation}
From the formulas reported in Table~\ref{tab:formulas}, one note that the dynamics is reversible in time in the absence of the non-conservative force; if $A=0$, then Eqs.~\eqref{eq:Brownian_Gyrator} describe two decoupled auto-regressive processes of order one, which are time-reversible in the steady state if $|B| < 1$. 

The logit can be easily derived following standard techniques able to deal with linear Gaussian systems in order to derive the stationary PDF $\mathcal{P}(\boldsymbol{x}|F)$; from here, one applies Bayes' Theorem and from the definition of the logit given by Eq.~\eqref{logistic} one obtains:
\begin{equation}
\label{eq:l_BG}
\ell(\boldsymbol{x}) = 2\frac{ x_{1}x_{4} - x_{2}x_{3}}{2B-B^2-A^2}.
\end{equation} 
In this case, $\ell(\boldsymbol{x})$ is the discrete-time version of the instantaneous work done by the non-conservative force on the particle~\cite{tome2006entropy}. In line with the second principle of thermodynamics, the more irreversible the dynamics in the steady state is, the more work is required to sustain the non-equilibrium stationary state. 
Note that $\ell(\boldsymbol{x})$ is symmetric about rotations of the phase space, but it increases in the radial direction: this is a clear signature of the coupling between the two coordinates if $A \neq 0$.

LR cannot describe $\ell(\boldsymbol{x})$ in Eq.~\eqref{eq:l_BG} since it has two contributions of second order in the encoded variables. 
\begin{figure}[tbh]
    \includegraphics[scale=0.55]{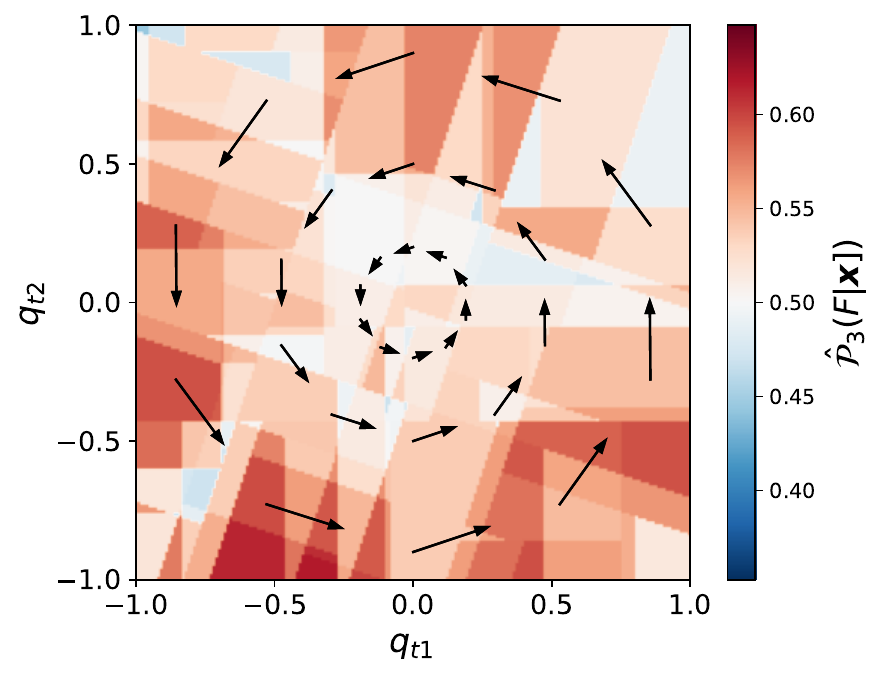}
    \includegraphics[scale=0.55]{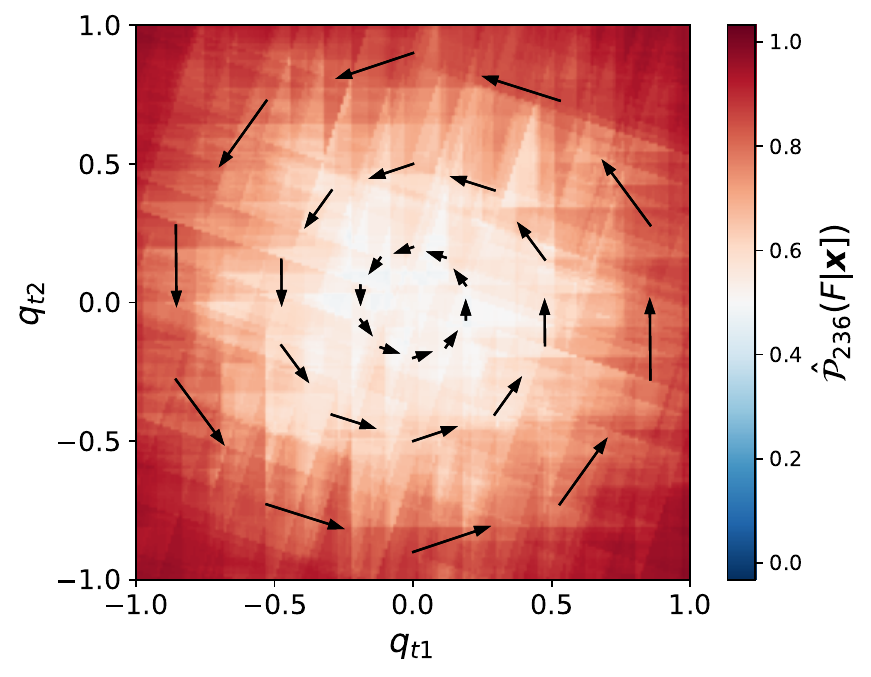}
    \caption{Estimated probability with XGBoost ($d_{max}=6$) from the Brownian gyrator with $A=0.3$ and $K=0.1$ for $N=100,000$ trajectories with an $80\%-20\%$ train-test split. Given that the true probability is 4-dimensional, the two dimensions not displayed in the plot are set as $E[\boldsymbol{q}_{t+1}|\boldsymbol{q}_t]$. Such expected jumps are represented by the black arrows in the plot. The upper panel shows the resulting coarse-graning with $N_{\text{tree}}=3$, while the lower panel after the early stopping activates, i.e.,  $N_{\text{tree}}=236$. }
    \label{fig:GB-coarse}
\end{figure}

For the sake of simplicity, we can say that the classifier, to guess the temporal direction correctly, should be able to identify a quadrant, so a simultaneous combination of the two input features. GB easily handles these cases by partitioning the feature space into trees of depth $d\geq2$ and sequentially combining them.

To better understand how GB estimates $\mathcal{P}(F|\boldsymbol{x})$, we present in Fig.~\ref{fig:GB-coarse} a two-dimensional projection of the non-homogeneous coarse-graining identified by the calibrated model. Specifically, given $\boldsymbol{q}_{t}=\{q_{t1},q_{t2}\}$ the contour plot shows the $\hat{\mathcal{P}}(F|\boldsymbol{x})$ by setting the last two features as $\{\overrightarrow x_3, \overrightarrow x_4\}= E[\boldsymbol{q}_{t+1}|\boldsymbol{q}_t]$, where the expectations are trivially derivable from Eq.~\eqref{eq:Brownian_Gyrator}. Similarly to the previous cases, the points around $\{q_{t1},q_{t1}\}=\{0,0\}$ have the largest sample size, for this reason, they have finer patches, whereas the points far from the origin are rarer and the related patches are wider. It is worth noting how the rotational symmetry emerges as the patches accumulate through boosting. This can be observed by tracking the resulting $\hat{\mathcal{P}}(F|\boldsymbol{x})$ from a small number of trees, as depicted in the upper panel of Fig.~\ref{fig:GB-coarse}, to the one resulting by boosting a high number of trees, as shown in the lower panel.

Whereas a model calibrated with $d \geq 2$ is able to correctly guess the shape of $\mathcal{P}(F|\boldsymbol{x})$, if the maximum depth is constrained to $d = 1$, then GB is no longer able to reconstruct the irreversibility, as shown by the points distributed on the flat line $\hat{\mathcal{P}}(F|\boldsymbol{x})=0.5$ in Fig.~\ref{fig:gauss_cauchy}. Note that the slight deviation of the points from $0.5$ is attributed to some degree of overfitting, which cannot be completely resolved even with early stopping (see Sec.~\ref{sec:pseudo}).

If we use a more sophisticated/optimal encoding that incorporates the coupling structure, such as
\begin{equation}\label{eq:optimalenc}
 \overrightarrow{\boldsymbol{x}}^*=\{ \overrightarrow{x}_1^*, \overrightarrow{x}_2^*\} = \{ q_{t1}q_{t+1,2},q_{t2}q_{t+1,1}\},    
\end{equation}
then a maximum depth $d=1$ is enough for an accurate estimation. This approach, in general, should be preferred since it effectively reduces the complexity and the potential for overfitting, as shown by the gray dots in the lower panel of Fig.~\ref{fig:gauss_cauchy}. This example explains how we can use a combination of different encodings and constrain the maximal depth of GB to guess the functional decomposition of the irreversibility. 

To summarize, when using a max depth of $1$, the model does not detect any irreversibility, meaning that no uncoupled terms are included in \( \ell(\boldsymbol{x}) \). However, testing some combinations of couplings can reveal an optimal encoding $\boldsymbol{x}^*$ that not only retrieves the same information as $\boldsymbol{x}$ but also improves estimation by reducing overfitting.

\section{Financial application}
\label{sec:finance}
In the following analysis, we aim to evaluate irreversibility in financial data using our method. Financial time series exhibit asymmetry with respect to time reversal transformations~\cite{zumbach2009time}, making them an ideal case study for validating our approach. Time irreversibility is a feature that should be expected according to some of the most popular models, such as the GARCH family~\cite{chicheportiche2014fine}, although some others, such as all continuous-time stochastic volatility models, do not predict such a feature~\cite{blanc2017quadratic}. Empirically, time irreversibility is observed in several works. The two most well-known related stylized facts are the so-called Zumbach \cite{zumbach2009time, zumbach2009time} and leverage effect \cite{borland}; the former implies that the correlation between large-scale past volatilities and small-scale future volatilities is larger than between small-scale past volatilities and large-scale future volatilities, while the latter implies that volatility tends to increase more after a price drop than after a price hike. Interestingly, time irreversibility, on top of representing a strong constraint for realistic stylized models of market dynamics, provides a piece of evidence against the efficient market dogma since it captures a direct impact of past price changes on future decisions and behavior of traders~\cite{borland}.

\subsection{Data and Preprocessing}
For our analysis, we use daily adjusted close-to-close returns from the ``WIKI Prices'' database \cite{QuandlWIKI}. WIKI Prices is a database collected from Quandl, a data aggregator that is now part of Nasdaq. This dataset is free, and it consists of $3199$ American equities from the NYSE and Nasdaq from a period spanning from 02-01-1962 to 11-04-2018, from which we retain only data after 1990. This dataset includes delisted stocks. To our knowledge, it is the only publicly available survivor bias-free dataset.

With the aim of measuring the irreversibility of a generic price trajectory, we ignore stock-specific details by considering all the trajectories at our disposals as coming from the same statistical ensemble. Therefore, irreversibility is averaged across all the stocks. 

We start by encoding raw stock prices into logarithmic returns, a common practice in financial datasets to achieve stationarity, i.e. 
\begin{equation}
    \overrightarrow{x}_i = \log(q_{i+1})-\log(q_{i}).
\end{equation}
As pointed out in Sec.~\ref{sec:encoding}, reversing the price trajectory in this encoding implies  
\begin{equation}
    \overleftarrow{\boldsymbol{x}}=\{-\overrightarrow{x}_{k}, \cdots, -\overrightarrow{x}_{1}\}
\end{equation}

However, given that stocks in the same time window are typically strongly correlated to each other, we must avoid having stocks on the same time window in both the train and test sets so to avoid over-fitting. 
Therefore, we use a $10$-Fold CV arranged by days rather than by stock. 
The maximum depth of the trees is set to $d=7$, the learning rate to $\lambda=0.3$, and the stopping rule on 10 iterations.

\begin{figure}[tbh]
  \centering

 \includegraphics[scale = 0.42]{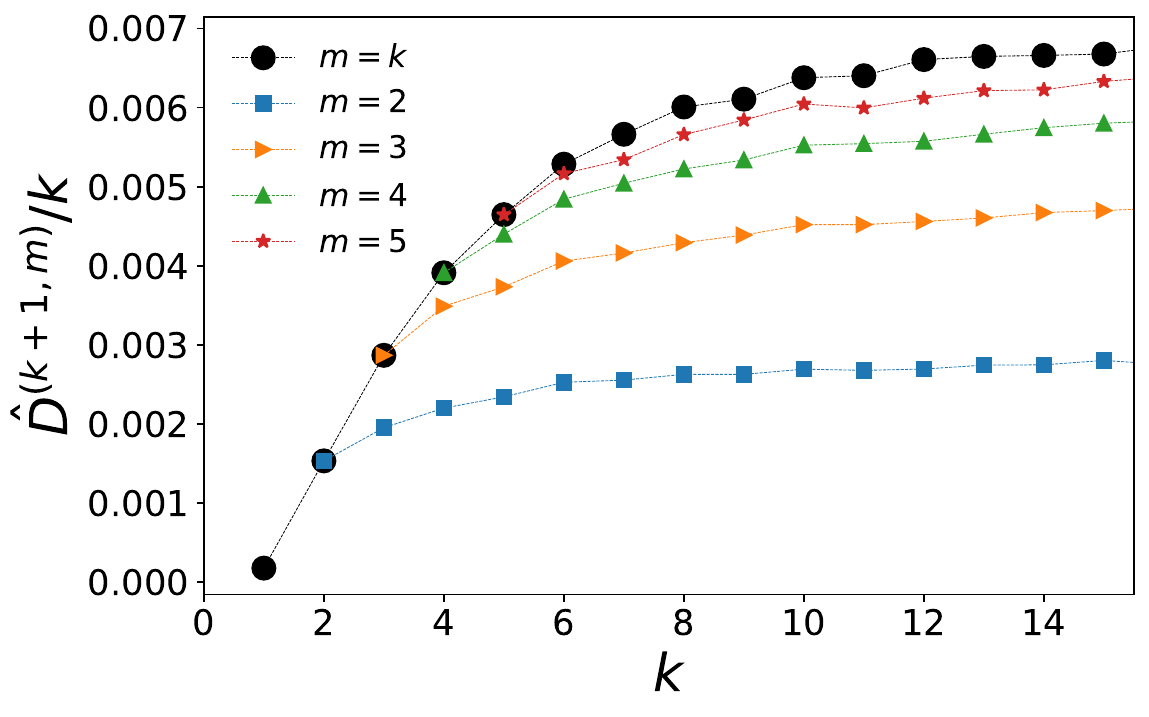} 
 
 \includegraphics[scale = 0.55]{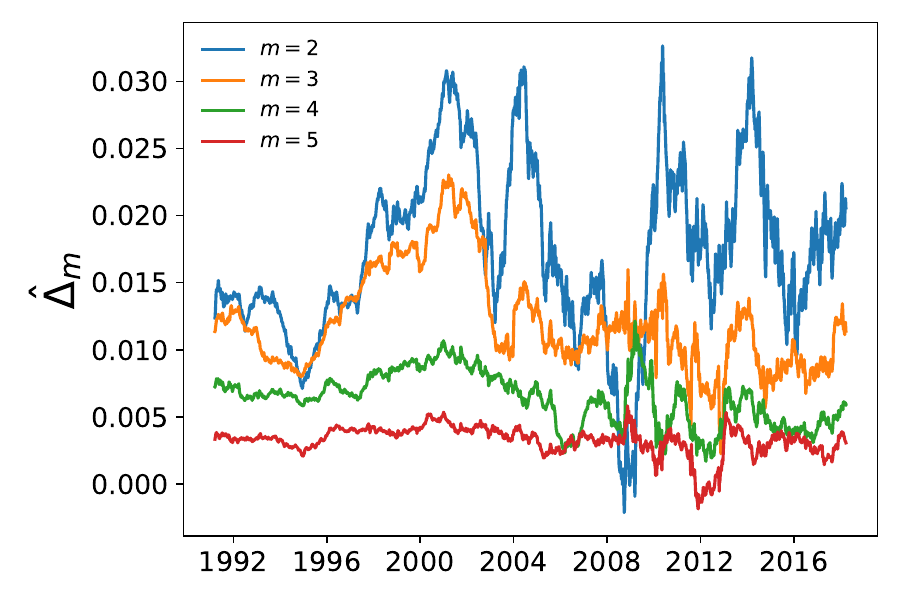}
    \caption{Analysis of the financial dataset. The top panel shows the average irreversibility per day as a function of the number of daily returns $k$ on the full historical period. Different colors refer to different interaction ranges $m$.  The bottom panel shows the day-dependent contribution to the irreversibility for specific interaction ranges $m$ and fixed $k = 7$.  Every line is smoothed with a moving average of 300 days. }
\label{fig:finance}

\end{figure}

\subsection{Analysis of the Markovian Time Range}
\label{sec:D_k}
First, we study $D^{(t)}$ to identify the order of Markovianity as explained in Sec.~\ref{sec:length}. An interesting initial observation emerges: $D^{(2)}$ is almost zero, as shown in the top plot of Fig.~\ref{fig:finance} (see black dots). This result is expected since we are focusing on $t=2$ prices, i.e., $k=1$ return and the daily return distribution is known to be centered at zero and almost symmetric. 

The black line in Fig.~\ref{fig:finance} shows that $\hat{D}^{(k+1)}/k$ reaches a plateau after $k>7$; this means that at $k=7$ we are sure to be already in the linear regime of $D^{(k+1)}$ (see Sec.~\ref{sec:length}), by setting our time-window $t=8$ we hope to have included all the detectable mechanisms contributing to irreversibility.

Note that the increments of irreversibility decrease for larger lags, meaning that short-term patterns contribute the most.

\subsection{Inspecting Irreversibility Functional Components}

The analysis we performed in the previous section with the unconstrained GB (Sec.~\ref{sec:D_k}) has revealed that key contributions to irreversibility are included within $7$ returns on average over the full historical period (1990-2018). At present, nothing is known about the presence of temporal variability in the system and, if so, how this impacts the coupling structure's behavior.

The unconstrained case can consider all possible coupling combinations among the encoded variables of any order, limited from above by the maximum depth which is set to $d = 7$. Now, we dissect the irreversibility to shed light on the individual contributions of these terms to simplify and identify their underlying functional components. We start from a simple ansatz: only the coupling among features related to two consecutive days is allowed i.e., $\{\{x_0,x_1\},\{x_1,x_2\},\cdots, \{x_{k-1},x_{k}\}\}$.
Such an analysis can be carried out using the ICs of XGBoost. Given that each chunk of size $m= 2$ is treated independently by the algorithm, this modification allows to capture periodicities of size $m$ in a window of length $k$. This procedure can be generalized to all combinations of $m$ consecutive days, where $m=k$ coincides with the unconstrained case. This discussion shows how interaction constraints allow us to add another layer to the analysis of irreversibility's functional form. We therefore extend the notation of the irreversibility to $D^{(t,m)}$ to include the interaction order $m$.

As a consistency check, note that the plateau of $\hat{D}^{(k+1,m)}/k$  for large $k$ depends upon the specific interaction constraint (displayed by the colored lines in the upper panel of Fig.~\ref{fig:finance}). In particular, it increases as $m$ grows, as expected. Moreover, for $m\geq5$, the maximum value approximates the prediction of an unconstrained classifier. This suggests that beyond a certain threshold of complexity ($m = 5$), adding more interactions does not significantly enhance the model's performance, enabling therefore a description of the time series with lower-order interactions~\cite{roldan2010estimating}. 

However, this analysis is day-independent, and financial markets are known to be strongly dependent on external conditions. As described in Sec.~\ref{sec:pseudo}, the analysis can be stratified according to the day. With a little more effort, one can identify the specific time-dependent contribution of different interaction orders. To do that, we define, for fixed $t=k+1$,
\begin{equation}
\label{eq:interactino_constr}
    \Delta_m= D^{(t,m)}-D^{(t,m-1)},
\end{equation}
that is the increase of irreversibility extending the interaction range by one at a given day. In doing this we are effectively isolating the contribution to irreversibility of interactions operating at different timescales.

The results of this investigation are in the bottom panel of Fig.~\ref{fig:finance}. While day-to-day interactions ($m = 2$) exhibit high variability, higher-order interaction $m>2$ contributions are more stable. 
Interestingly, during periods of crisis, the irreversibility with constraint \(m=2\) drops to 0 and even below; the latter case indicates that, on average, our procedure misclassifies price trajectories more often than a random guess. This effect is particularly relevant around the global financial crisis of 2008.

This phenomenon unveils significant insights into the Markovian nature of the trajectory: it suggests that typical short-term patterns are disrupted during turbulent market phases.
In these circumstances, only robust long-term interactions are left, which unfortunately are more difficult to model and estimate with traditional statistical methods.

\section{Discussion}
\label{sec:5}  
We propose a gradient boosting approach for estimating irreversibility in multivariate time series by leveraging a well-known mapping onto a classification task. Through gradient boosting, we harnessed a model-free, nonlinear estimation requiring minimal calibration of the classifier. We present alternatives to gradient boosting and thoroughly discuss their regime of applicability, highlighting the advantages of the proposed technique. 

An additional functionality of the proposed methodology is
to easily dissect the contributions to the irreversibility of subsets of variable interactions, for instance,
those operating at different time scales. The procedure we propose can be schematized into three phases. The first one involves identifying an appropriate encoding of the trajectory. By selecting an insightful encoding, one can improve the accuracy of the irreversibility estimation. For example, if prior knowledge about the system suggests certain dynamical symmetries one can remove non-informative parts for what concerns irreversibility estimation by considering appropriate variable combinations, leading to a more efficient estimator. Alternatively, if the data-generating model is unknown, different encodings can be used to test ad-hoc hypotheses. The second phase focuses on identifying the Markovian order of the trajectory using gradient boosting. This step helps determine the minimal length of trajectory slices needed to describe irreversibility completely. The final phase involves hypothesis testing to uncover interactions within the underlying model. One can test the presence of variables' interactions by employing interaction constraints within the gradient boosting framework. This step allows for a detailed decomposition of irreversibility in its functional components, revealing, for instance, how different interactions operating at different timescales contribute to the overall irreversibility.

Our methodology reveals interesting insights when applied to one of the most challenging and complex systems, i.e., financial markets. Using our method, we estimate the order of Markovianity to be smaller than two weeks. Then, we decompose irreversibility in contributions, to discover how events at different time scales interact. By analyzing the time dependence of these contributions, we spot a distinctive shift from a predominantly short interaction range during stable periods to significantly higher ones during financial crises. This finding highlights the crucial role of long-range time interactions in studying financial markets during turbulent times.

\vspace{0.5cm}

\section*{Author Contributions}
M.V.~and C.B.~conceptualized the problem and wrote the manuscript. C.B.~supervised the project and proposed the methodology. C.P.~performed the analysis and wrote the first version of the draft. M.V.~integrated the methodology into the research framework. All authors contributed to the discussion and interpretation of the results.

\section*{Data Availability Statement}
The Python library ``time\_irreversibility\_estimator'' is available on PyPI. This library implements the calculation of irreversibility using the K-Fold approach with XGBoost. The library's detailed documentation and usage instructions can be accessed via its GitHub repository \cite{irreversibility_estimator_2024}.
The financial data used in this study are freely available for download at Ref.~\cite{QuandlWIKI}.

\section*{Acknowledgments}
This work was performed using HPC resources from the ``Mésocentre'' computing center of CentraleSup\'elec and \'Ecole Normale Sup\'erieure Paris-Saclay supported by CNRS and R\'egion \^Ile-de-France (\url{http://mesocentre.centralesupelec.fr/}). M.V. is supported by the Agence National de la Recherche (CogFinAgent: ANR-21-CE23-0002-02).
\bibliographystyle{apsrev4-2}
\bibliography{biblio}

\vspace{1cm}

\end{document}